\documentclass[pre,reprint,superscriptaddress]{revtex4-1} 

\usepackage{blkarray}
\usepackage{amsmath,amssymb}
\usepackage{graphicx}
\usepackage{hyperref}

\newcommand{\beq}{\begin{equation}}
\newcommand{\eeq}{\end{equation}}

\begin{document}

\title{Integrated Information in the Spiking-Bursting Stochastic Model}

\author{Oleg Kanakov}
\author{Susanna Gordleeva}
\affiliation{Lobachevsky State University of Nizhny Novgorod, Nizhny Novgorod, 
	Russia}	

\author{Alexey Zaikin}\email[E-mail: ]{alexey.zaikin@ucl.ac.uk}
\affiliation{Lobachevsky State University of Nizhny Novgorod, Nizhny Novgorod, 
	Russia}	
\affiliation{Institute for Women's Health and Department of Mathematics, 
	University College London, London, United Kingdom}
\affiliation{Department of Pediatrics, Faculty of Pediatrics, Sechenov University, Moscow, Russia}

\begin{abstract}
	This study presents a comprehensive analytic description in terms of the empirical ``whole minus sum'' version of Integrated Information in comparison to the ``decoder based'' version for the ``spiking-bursting'' discrete-time, discrete-state stochastic model, which was recently introduced to describe a specific type of dynamics in a neuron-astrocyte network. The ``whole minus sum'' information may change sign, and an interpretation of this transition in terms of ``net synergy'' is available in the literature. This motivates our particular interest to the sign of the ``whole minus sum'' information in our analytical consideration. The behavior of the ``whole minus sum'' and ``decoder based'' information measures are found to bear a lot of similarity, showing their mutual asymptotic convergence as time-uncorrelated activity is increased, with the sign transition of the ``whole minus sum'' information associated to a rapid growth in the ``decoder based'' information. The study aims at creating a theoretical base for using the spiking-bursting model as a well understood reference point for applying Integrated Information concepts to systems exhibiting similar bursting behavior (in particular, to neuron-astrocyte networks). The model can also be of interest as a new discrete-state test bench for different formulations of Integrated Information.
\end{abstract}

\maketitle

\section{Introduction}
Integrated information (II) \cite{tononi2004information, balduzzi2008integrated, tononi2012integrated, oizumi2014phenomenology} is a measure of internal information exchange in complex systems which was initially proposed to quantify consciousness \cite{tononi2008consciousness}. This initial aim still remaining a matter of research and debate \cite{peressini2013consciousness, tsuchiya2016using, tononi2016integrated, norman2017quantum}, the II concept itself is by now a widely acknowledged tool in the field of complex dynamics analysis \cite{engel2018integrated, 2016arXiv160608313M, toker2019information}. The general concept gave rise to specific ``empirical'' formalizations of II \cite{barrett2011practical, 2014arXiv14010978G, oizumi2016unified, oizumi2016measuring} aimed at computability from empirical probability distributions based on real data. For a systematic taxonomy of II measures see \cite{tegmark2016improved}, and a comparative study of empirical II measures in application to Gaussian autoregressive network models has been recently done in \cite{mediano2019measuring}.

A recent study \cite{AstroNeur} addressed the role of astrocytic regulation of neurotransmission \cite{araque2014} in generating positive II by small networks of brain cells --- neurons and astrocytes. Empirical ``whole minus sum'' II as defined in \cite{barrett2011practical} was calculated in \cite{AstroNeur} from the time series produced by a biologically realistic model of neuro-astrocytic networks. A simplified, analytically tractable stochastic ``spiking-bursting'' model (in complement to the realistic one) was designed to describe a specific type of activity in neuro-astrocytic networks which manifests itself as a sequence of intermittent system-wide excitations of rapid pulse trains (``bursts'') on the background of random ``spiking'' activity in the network. The spiking-bursting model is a discrete-time, discrete-state stochastic process which mimics the main features of this behavior. The model was successfully used in \cite{AstroNeur} to produce semi-analytical estimates of II in good agreement with direct computation of II from time series of the biologically realistic network model.

The present study aims at creating a theoretical base for using the spiking-bursting model of \cite{AstroNeur} as a well understood reference point for applying Integrated Information concepts to systems exhibiting similar bursting behavior (in particular, to other neuron-astrocyte networks). We also aim at extending the knowledge of comparative features of different empirical II measures, which are currently available mainly in application to Gaussian autoregressive models \cite{mediano2019measuring, tegmark2016improved}, by applying two such measures \cite{barrett2011practical, oizumi2016measuring} to our discrete-state model.

In Sections~\ref{sec_defs}, \ref{sec_model} we specify the definitons of the used II measures and the model. Specific properties of the model which lead to a redundance in its parameter set are addressed in Section~\ref{sec_scaling}.
In Section~\ref{sec_ii} we provide with an analytical treatment for the empirical ``whole minus sum'' \cite{barrett2011practical} version of II in application to our model. This choice among other empirical II measures is inherited from the preceding study \cite{AstroNeur} and is in part due to its easy analytical tractability, and also due to its ability to change sign, which naturally identifies a transition point in the parameter space. This property may be considered a violation of the natural non-negativeness requirement for II \cite{oizumi2016measuring}; on the other hand, the sign of the ``whole minus sum'' information has been given interpretation in terms of ``net synergy'' \cite{barrett2015exploration} as a degree of redundancy in the evolution of a system \cite{mediano2019measuring}. In this sense this transition may be viewed as a useful marker in its own right in the toolset of measures for complex dynamics. This motivates our particular focus on identifying the sign transition of the ``whole minus sum'' information in the parameter space of the model. We also identify a scaling of II with a parameter determining time correlations of the bursting (astrocytic) subsystem when these correlations are weak.

In Section~\ref{sec_cmp} we compare the outcome of the ``whole minus sum'' II measure \cite{barrett2011practical} to the ``decoder based'' measure $\Phi^{\ast}$, which was specifically designed in \cite{oizumi2016measuring} to satisfy the non-negativeness property. We compute $\Phi^{\ast}$ directly by definition from known probability distributions of the model. Despite their inherent difference consisting in changing or not changing sign, the two compared measures are shown to bear similarities in their dependence upon model parameters, including the same scaling with the time correlation parameter.


\section{Definition of II Measures in Use}\label{sec_defs}
The empirical ``whole minus sum'' version of II is formulated according to
\cite{barrett2011practical} as follows. Consider a stationary
stochastic process $\xi(t)$ (binary vector process), whose
instantaneous state is described by $N$ binary digits (bits), each identified with a node of the network (neuron). The
full set of $N$ nodes (``system'') can be split into two
non-overlapping non-empty subsets (``subsystems'') $A$ and
$B$, such splitting further referred to as bipartition $AB$.
Denote by $x=\xi(t)$ and $y=\xi(t+\tau)$ two states of the process
separated by a specified time interval $\tau\neq0$. States of the
subsystems are denoted as $x_A$, $x_B$, $y_A$, $y_B$.

Mutual information between $x$ and $y$ is defined as
\beq\label{eq_defIxy}
I_{xy}=H_x+H_y-H_{xy},
\eeq
where
\beq\label{eq_Hx}
H_x=-\sum_x p(x) \log_2 p(x)
\eeq
is entropy (base 2 logarithm gives
result in bits), summation is hereinafter assumed to be taken over the whole range of the index variable (here $x$), $H_y=H_x$ due to assumed stationarity.

Next, a
bipartition $AB$ is considered, and ``effective information'' as a
function of the particular bipartition is defined as
\beq
\label{eq_Ieff}
\Phi_{\text{eff}}(AB)=I_{xy}-I_{x_A,y_A}-I_{x_B,y_B}.
\eeq

II is then defined as effective information
calculated for a specific bipartition $AB^{\text{MIB}}$ (``minimum
information bipartition'') which minimizes specifically normalized
effective information:
\begin{subequations}\label{eq_II}
	\begin{gather}
		\Phi=\Phi_{\text{eff}}(AB^{\text{MIB}}),\\		
		AB^{\text{MIB}}=\mathrm{argmin}_{AB} 
		\left[\frac{\Phi_{\text{eff}}(AB)}{\min\{H(x_A),H(x_B)\}} \right].\label{eq_II_b}
	\end{gather}
\end{subequations}
Note that this definition prohibits positive II, whenever $\Phi_{\text{eff}}$ turns out to be zero or negative for at least one bipartition $AB$.

We compare the result of the ``whole minus sum'' effective information \eqref{eq_Ieff} to the ``decoder based'' information measure $\Phi^{\ast}$, which is modified from its original formulation of \cite{oizumi2016measuring} by setting the logarithms base to 2 for consistency:
\begin{subequations}\label{eq_Phistar}
\beq
\Phi^{\ast}(AB) = I_{xy} - I^{\ast}_{xy}(AB),
\eeq
where
\begin{multline}
I^{\ast}_{xy}(AB) = \max_{\beta} \left[ -\sum_y p(y) \log_2 \sum_x p(x) q_{AB}(y|x)^{\beta}\right.\\
\left. + \sum_{xy} p(xy) \log_2  q_{AB}(y|x)^{\beta} \right],
\end{multline}
\begin{equation}
	q_{AB}(y|x) = p(y_A|x_A) p(y_B|x_B) = \frac{p(x_A y_A) p(x_B y_B)}{p(x_A) p(x_B)}.
\end{equation}
\end{subequations}

\section{Spiking-Bursting Stochastic Model}\label{sec_model}

We consider a stochastic model, which produces a binary vector valued, discrete-time stochastic process.
In keeping with \cite{AstroNeur}, this ``spiking-bursting'' model is defined as a combination $M=\{V, S\}$ of a time-correlated 
dichotomous component $V$ which turns on and off system-wide bursting (that mimics global bursting of a neuronal network, when each neuron produces a train of pulses at a high rate \cite{AstroNeur}), and a 
time-uncorrelated component $S$ describing spontaneous (spiking) activity (corresponding to a background random activity in a neural network characterized by relatively sparse random appearance of neuronal pulses --- spikes \cite{AstroNeur}) occurring in the 
absence of a burst. The model mimics the spiking-bursting type of activity which occurs in a neuro-astrocytic network, where the neural subsystem normally exhibits time-uncorrelated patterns of spiking activity, and all neurons are under the common influence of the astrocytic subsystem, which is modeled by the dichotomous component $V$ and sporadically induces simultaneous bursting in all neurons. A similar network architecture with a ``master node'' spreading its influence on subordinated nodes was considered for example in \cite{tononi2004information} (Figure~4b therein).

The model is defined as follows. At each instance of (discrete) time the state of the dichotomous component can be either ``bursting'' with probability $p_b$, or ``spontaneous'' (or
``spiking'') with probability $p_s=1-p_b$. While in the bursting mode,
the instantaneous state of the resulting process $x=\xi(t)$ is given by 
all ones: $x=11..1$ 
(further abbreviated as $x=1$). In case of spiking, the state $x$ is a (time-uncorrelated) random variate described by a discrete probability distribution $s_x$ (where an occurrence of `1' in any bit is referred to as a ``spike''), so that the resulting one-time state probabilities read
\begin{subequations}\label{eq_Ponetime}
	\begin{align}
		p(x\neq1) &= p_s s_x,\\
		p(x=1)    &= p_1, \quad p_1=p_s s_1 + p_b,\label{eq_Ponetime_p1}
	\end{align}
\end{subequations}
where $s_1$ is the probability of spontaneous occurrence of $x=1$ (hereafter referred to as a system-wide simultaneous \footnote{In a real network ``simultaneous'' implies occuring within the same time discretization interval \cite{AstroNeur}.} spike) in the 
absence of a burst.

To describe two-time joint probabilities for $x=\xi(t)$ and $y=\xi(t+\tau)$, consider a joint
state $xy$ which is a concatenation of bits in $x$ and $y$. The
spontaneous activity is assumed to be uncorrelated in time, which leads to the factorization
\beq\label{eq_sxy_factor}
s_{xy}=s_x s_y.
\eeq
The time correlations of the dichotomous component \footnote{In a neural network these correlations are conditioned by burst duration \cite{AstroNeur}; e.g., if this (in general, random) duration mostly exceeds $\tau$, then the correlation is positive.} are described by a $2\times 2$ matrix
\beq\label{eq_psbmat}
p_{q\in\{s, b \},r\in\{s, b \}}=
\begin{pmatrix}
	p_{ss} & p_{sb} \\
	p_{bs} & p_{bb}
\end{pmatrix}
\eeq
whose components are joint
probabilities to observe the respective spiking (index ``$s$'') and/or bursting (index ``$b$'') states in $x$
and $y$. The probabilities obey $p_{sb}=p_{bs}$ (due to
stationarity), $p_b=p_{bb}+p_{sb}$, $p_s=p_{ss}+p_{sb}$, thereby allowing to 
express all one- and two-time probabilities describing the dichotomous 
component in terms of two independent quantities, which for example can be a pair $\{p_s, p_{ss}\}$, then
\begin{subequations}\label{eq_pspss}
	\begin{align}
		p_{sb} &= p_{bs} = p_s - p_{ss},\\
		p_{bb} &= 1-(p_{ss} + 2 p_{sb}),
	\end{align}
\end{subequations}
or $\{p_b, \rho\}$ as in \cite{AstroNeur}, where $\rho$ is correlation coefficient defined by
\beq\label{eq_phi}
p_{sb}=p_s p_b (1-\rho).
\eeq
In Section~\ref{sec_scaling} we justify the use of another effective parameter $\epsilon$ \eqref{eq_ep} instead of $\rho$ to determine time correlations in the dichotomous component.

The two-time joint probabilities for the resulting process are then expressed as
\begin{subequations}\label{eq_Ptwotime}
	\begin{gather}
		\begin{align}
			p(x\neq 1, y\neq 1) &= p_{ss} s_x s_y, \\
			p(x\neq 1, y=1) &= \pi s_x, \quad p(x=1, y\neq 1) = \pi s_y, \\
			p(x=1, y=1) &= p_{11},
		\end{align}\\
		\pi=p_{ss} s_1 + p_{sb}, \quad p_{11}=p_{ss}
		s_1^2 + 2 p_{sb} s_1 + p_{bb}.\label{eq_Ptwotime_p11}
	\end{gather}
\end{subequations}

Note that the above notations can be applied to any subsystem
instead of the whole system (with the same dichotomous component, as it is 
system-wide anyway).

\section{Model Parameters Scaling}\label{sec_scaling}
The spiking-bursting stochastic model as described in Section~\ref{sec_model} is redundant in the following sense. In terms of the model definition, there are two distinct states of the model which equally lead to observing the same one-time state of the resultant process with 1's in all bits: firstly --- a burst, and secondly --- a system-wide simultaneous spike in the absence of a burst, which are indistinguishable by one-time observations. Two-time observations reveal a difference between system-wide spikes on one hand and bursts on the other, because the latter are assumed to be correlated in time, unlike the former. That said, the ``labeling'' of bursts versus system-wide spikes exists in the model (by the state of the dichotomous component), but not in the realizations. Proceeding from the realizations, it must be possible to relabel a certain fraction of system-wide spikes into bursts (more precisely, into a time-uncorrelated portion thereof). Such relabeling would change both components of the model $\{V,S\}$ (dichotomous and spiking processes), in particular diluting the time correlations of bursts, without changing the actual realizations of the resultant process. This implies the existence of a transformation of model parameters which keeps realizations (i.e. the stochastic process as such) invariant. The derivation of this transformation is presented in Appendix~\ref{sec_scaleder} and leads to the following scaling
\begin{subequations}\label{eq_scaling}
	\begin{align}
		s_x &= \alpha s'_x, \\
		1-s_1 &= \alpha (1- s'_1),\\
		p_{s'}&=\alpha p_s,\\
		p_{s's'} &= \alpha^2 p_{ss},
	\end{align}
\end{subequations}
where $\alpha$ is a positive scaling parameter, and all other probabilities are updated according to Eq.~\eqref{eq_pspss}.

The mentioned invariance in particular implies that any characteristic of the process must be invariant to the scaling (\ref{eq_scaling}a-d). This suggests a natural choice of a scaling-invariant effective parameter $\epsilon$ defined by
\beq\label{eq_ep}
p_{ss}=p_s^2 (1+\epsilon)
\eeq
to determine time correlations in the dichotomous component. In conjunction with a second independent parameter of the dichotomous process, for which a straightforward choice is $p_s$, and with full one-time probability table for spontaneous activity $s_x$, these constitute a natural full set of model parameters $\{s_x, p_s, \epsilon\}$.

The two-time probability table \eqref{eq_psbmat} can be expressed in terms of $p_s$ and $\epsilon$ by substituting Eq.~\eqref{eq_ep} into Eq.~\eqref{eq_pspss}:
\beq\label{eq_psbeps}
p_{q\in\{s, b \},r\in\{s, b \}}=\left(
\begin{array}{rr}
	p_s^2 +\epsilon p_s^2 & p_s p_b -\epsilon p_s^2 \\
	p_s p_b -\epsilon p_s^2 & p_b^2 +\epsilon p_s^2
\end{array}\right).
\eeq
The requirement of non-negativeness of probabilities imposes simultaneous constraints
\begin{subequations}\label{eq_psmax}
\begin{gather}
\epsilon\ge -1 \\
\intertext{and}
p_s\le p_{s \max}=
\begin{cases}
	\frac{1}{1+\epsilon}\left(1-\sqrt{|\epsilon|}\right) & \text{if} \quad -1\le \epsilon<0, \\
	\frac{1}{1+\epsilon} & \text{if} \quad \epsilon \ge 0,
\end{cases}
\end{gather}
\end{subequations}
or, equivalently,
\beq\label{eq_epmax}
-\epsilon_{\max}^2 \le \epsilon\le \epsilon_{\max}=\frac{p_b}{p_s}.
\eeq

Comparing the off-diagonal term $p_{sb}$ in \eqref{eq_psbeps} to the definition of correlation coefficient $\rho$ in \eqref{eq_phi}, we get
\beq\label{eq_phieps}
\epsilon = \rho\: \frac{p_b}{p_s}  = \rho\: \epsilon_{\max},
\eeq
thus the sign of $\epsilon$ has the same meaning as that of $\rho$. Hereinafter we limit ourselves to non-negative correlations $\epsilon\ge 0$. 

\section{Analysis of the Empirical ``Whole Minus Sum'' Measure for the Spiking-Bursting Process}\label{sec_ii}
In this Section we analyze the behavior of the ``whole minus sum'' empirical II \cite{barrett2011practical} defined by Eqs.~\eqref{eq_Ieff}, \eqref{eq_II} for the spiking-bursting model in dependence of the model parameters, particularly focusing on its transition from negative to positive values.

Mutual information $I_{xy}$ for two time instances $x$ and $y$ of the spiking-bursting process is expressed by inserting all one- and two-time probabilities of the process according to \eqref{eq_Ponetime}, \eqref{eq_Ptwotime} into the definition \eqref{eq_defIxy}, \eqref{eq_Hx}. The full derivation is given in Appendix~\ref{sec_deriv} and leads to an expression which was used in \cite{AstroNeur}
\begin{multline}\label{eq_Ixy}
	I_{xy}=2(1-s_1)\{p_s\} + 2\{p_1\} -(1-s_1)^2\{p_{ss}\}\\
	-2(1-s_1)\{\pi\} - \{p_{11}\}, 
\end{multline}
where we denote for compactness
\beq\label{eq_braces}
\{q\}=-q \log_2 q.
\eeq

We exclude from further consideration the following degenerate cases which automatically give $I_{xy}=0$ by definition \eqref{eq_defIxy}:
\beq\label{eq_degenerate}
s_1=1,\quad \text{or} \quad p_s=0,\quad \text{or} \quad p_s=1,\quad \text{or} \quad \rho=\epsilon=0,
\eeq
where the former two correspond to a deterministic ``always 1'' state for which all entropies in \eqref{eq_defIxy} are zero, and the latter two produce no predictability, which implies $H_{xy}=H_x+H_y$.

\begin{figure}
	\includegraphics[width=\columnwidth]{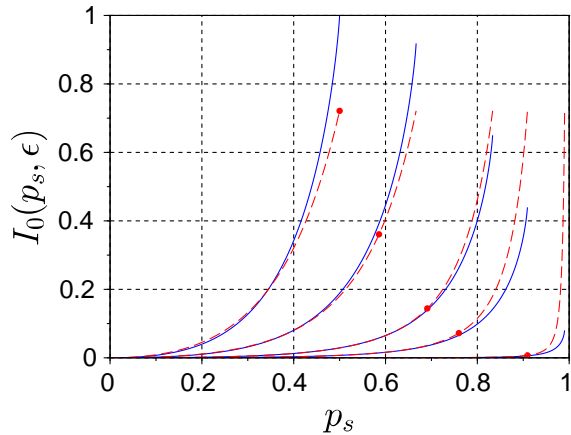}
	\caption{Blue solid lines --- plots of $I_0(p_s, \epsilon)$ versus $p_s$ varied from 0 to $p_{s\max}$ as per \eqref{eq_psmax}, at $\epsilon=0.01$, 0.1, 0.2, 0.5, 1 (from right to left). Red dashed lines --- approximation \eqref{eq_I0appx}. Red dots --- upper bounds of approximation applicability range \eqref{eq_I0appx_condc}.}\label{fig_I0}
\end{figure}

The particular case $s_1=0$ in \eqref{eq_Ixy} reduces to

\begin{equation}
\begin{split}\label{eq_I0}
I_{xy}|_{s_1=0} &= 2\big(\{p_s\} + \{p_b\}\big) - \big(\{p_{ss}\} + 2\{p_{sb}\} + \{p_{bb}\} \big)\\
& =I_0(p_s, \epsilon),
\end{split}
\end{equation}
which coincides with mutual information for the dichotomous component taken alone and can be seen as a function denoted in \eqref{eq_I0} as $I_0(\cdot, \cdot)$ of just two independent parameters of the dichotomous component, for which we chose $p_s$ and $\epsilon$ as described in Section~\ref{sec_scaling}. Typical plots of $I_0(p_s, \epsilon)$ versus $p_s$ at fixed $\epsilon$ are shown with blue solid lines in Fig.~\ref{fig_I0}.

The general case \eqref{eq_Ixy} can be recovered back from \eqref{eq_I0} by virtue of the scaling (\ref{eq_scaling}a-d), by assuming $s'_1=0$ in (\ref{eq_scaling}b) and substituting the corresponding scaled value $p_{s'}=(1-s_1)p_s$ as per (\ref{eq_scaling}c) in place of the first argument of function $I_0(p_{s'}, \epsilon)$ defined in \eqref{eq_I0}, while parameter $\epsilon$ remains invariant to the scaling. This produces a simplified expression
\beq\label{eq_IxyI0}
I_{xy} = I_0\big((1-s_1)p_s, \epsilon \big),
\eeq
which is still exactly equivalent to \eqref{eq_Ixy}. We emphasize that hereinafter expressions containing $I_0(\cdot,\cdot)$ like \eqref{eq_IxyI0}, \eqref{eq_Ieff0}, \eqref{eq_sufcond_b} etc. imply that all probabilities in \eqref{eq_I0} must be expressed in terms of $p_s$ and $\epsilon$, and $p_s$ in turn be accordingly substituted by the actual first argument of $I_0(\cdot,\cdot)$, e.g. by $(1-s_1)p_s$ in \eqref{eq_IxyI0}. The same applies when the approximate expression for $I_0(\cdot)$ \eqref{eq_I0appx} is used.

Given a bipartition $AB$ (see Section~\ref{sec_defs}), this result is applicable as well
to any subsystem $A$ ($B$), with $s_1$ replaced by $s_A$ ($s_B$) which denote 
the probability of a subsystem-wide simultaneous spike $x_A=1$ ($x_B=1$) in the absence of a burst, and
with same parameters of the dichotomous component (here $p_s$, $\epsilon$).
Then effective information \eqref{eq_Ieff} is expressed as
\beq\label{eq_Ieff0}
\Phi_{\text{eff}} = I_0\big((1-s_1)p_s, \epsilon \big) - I_0\big((1-s_A)p_s, \epsilon \big) - I_0\big((1-s_B)p_s, \epsilon \big).
\eeq

Hereafter in this section we assume the independence of spontaneous activity across the system, which implies
\beq\label{eq_sAB}
s_A s_B = s_1,
\eeq
then \eqref{eq_Ieff0} turns into
\begin{subequations}\label{eq_fsdef}
\begin{equation}\label{eq_fsdef_a}
\Phi_{\text{eff}} = f(s_A),
\end{equation}
where
\begin{multline}\label{eq_fsdef_b}
f(s)= I_0\big((1-s_1)p_s, \epsilon \big) - I_0\big((1-s)p_s, \epsilon \big)\\
 - I_0\big((1-s_1/s)p_s, \epsilon \big).
\end{multline}
\end{subequations}Note that the function $I_0(\cdot, \cdot)$ in \eqref{eq_I0} is defined only when the first argument is in the range $(0,1)$, thus the definition domain of $f(s)$ in \eqref{eq_fsdef_b} is
\beq\label{eq_fsdom}
s_1<s<1.
\eeq

According to \eqref{eq_II}, the necessary and sufficient condition for the ``whole minus sum'' empirical II be positive is the requirement that $\Phi_{\text{eff}}$ be positive for any bipartition $AB$. Due to \eqref{eq_fsdef}, this requirement can be written in the form
\beq\label{eq_condmin}
\min_{s\in \{s_A\}} f(s)>0,
\eeq
where $\{s_A\}$ is the set of $s_A$ values for all possible bipartitions $AB$ (if $A$ is any non-empty subsystem, then $s_A$ is defined as the probability of spontaneous occurrence of 1's in all bits in $A$ in the same instance of the discrete time).

Expanding the set of $s$ in \eqref{eq_condmin} to the whole definition domain of $f(s)$ \eqref{eq_fsdom} leads to a sufficient (generally, stronger) condition for positive II
\beq\label{eq_condminsuf}
\min_{s_1<s<1} f(s)>0.
\eeq

\begin{figure}
	\centering
	\includegraphics[width=\columnwidth]{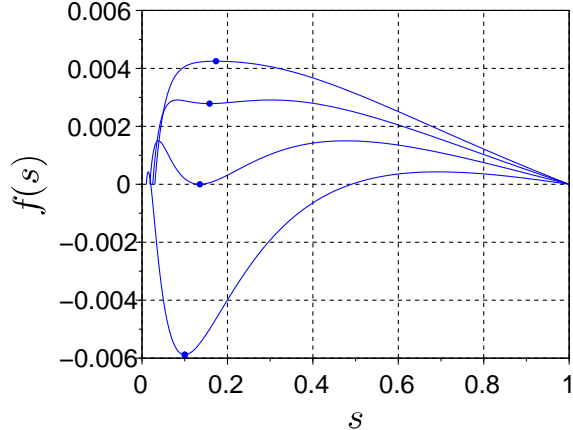}
	\caption{Plots of $f(s)$ on $s_1<s<1$ with $s_1=0.03$, 0.025, 0.0182919, 0.01 (from top to bottom) at $p_s=0.7$, $\epsilon=0.1$. For each value of $s_1$, the  extremum $(\sqrt{s_1}, f(\sqrt{s_1}))$ is indicated with a dot.}\label{fig_fs}
\end{figure}

Note \footnote{All mentioned properties and subsequent reasoning can be observed in Fig.~\ref{fig_fs}, which shows a few sample plots of $f(s)$.} that $f(s)$ by definition \eqref{eq_fsdef_b} satisfies $f(s=s_1)=f(s=1)=0$, $f'(s=s_1)>0$ and (due to the invariance to mutual renaming of subsystems $A$ and $B$) $f(s_1/s)=f(s)$. The latter symmetry implies that the quantity of extrema on $(s_1,1)$ must be odd, one of them always being at $s=\sqrt{s_1}$. If the latter is the only extremum, then it is a positive maximum, and \eqref{eq_condminsuf} is thus fulfilled automatically. In case of three extrema, $f(\sqrt{s_1})$ is a minimum, which can change sign. In both these cases the condition \eqref{eq_condminsuf} is equivalent to the requirement 
\beq\label{eq_sufcond0}
f(\sqrt{s_1})>0,
\eeq
which can be rewritten as
\begin{subequations}\label{eq_sufcond}
\beq\label{eq_sufcond_a}
g(s_1)>0,
\eeq
where
\beq\label{eq_sufcond_b}
g(s_1)=f(\sqrt{s_1})=I_0\big((1-s_1)p_s, \epsilon \big) - 2 I_0\big((1-\sqrt{s_1})p_s, \epsilon \big).
\eeq
\end{subequations}
The equivalence of \eqref{eq_sufcond0} to \eqref{eq_condminsuf} would be broken in case of 5 or more extrema. As suggested by numerical evidence \footnote{The equivalence of \eqref{eq_condminsuf} to \eqref{eq_sufcond} was confirmed up to machine precision for each combination of $p_s\in[0.01, 0.99]$ and $\rho\in[0.01, 1]$ (both with step 0.01).}, this exception never holds, although we did not prove this rigorously. Based on the reasoning above, in the following we assume the equivalence of \eqref{eq_sufcond0} (and \eqref{eq_sufcond}) to \eqref{eq_condminsuf}.

A typical scenario of transformations of $f(s)$ with the change of $s_1$ is shown in Fig.~\ref{fig_fs}. Here the extremum $f(\sqrt{s_1})$ (shown with a dot) transforms with the decrease of $s_1$ from a positive maximum into a minimum, which in turn decreases from positive through zero into negative values.

Note that by construction, the function $g(s_1)$ defined in \eqref{eq_sufcond_b} expresses effective information $\Phi_{\text{eff}}$ from \eqref{eq_Ieff} for a specific bipartition characterized by $s_A=s_B=\sqrt{s_1}$. If such ``symmetric'' bipartition exists, then the value $\sqrt{s_1}$ belongs to the set $\{s_A\}$ in \eqref{eq_condmin}, which implies that \eqref{eq_sufcond0} (same as \eqref{eq_sufcond}) is equivalent not only to \eqref{eq_condminsuf}, but also to the necessary and sufficient condition \eqref{eq_condmin}. Otherwise, \eqref{eq_condminsuf} (equivalently, \eqref{eq_sufcond0} or \eqref{eq_sufcond}), formally being only sufficient, still may produce a good estimate of the necessary and sufficient condition in cases when $\{s_A\}$ contains values which are close to $\sqrt{s_1}$ (corresponding to nearly symmetric partitions, if such exist).

\begin{figure*}
	\centering
	\textbf{(a)}\includegraphics[width=0.46\textwidth]{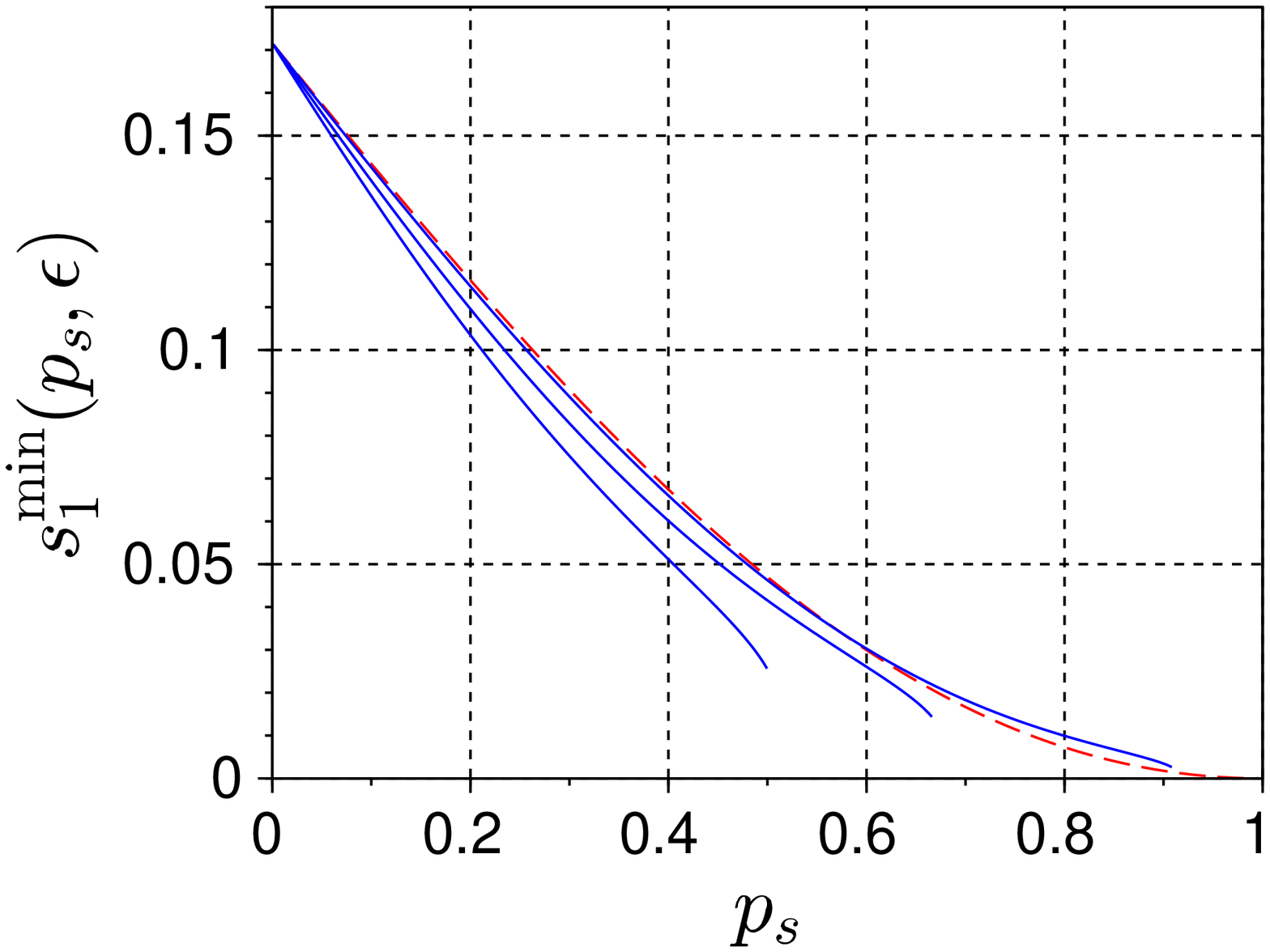}\hfill
	\textbf{(b)}\includegraphics[width=0.46\textwidth]{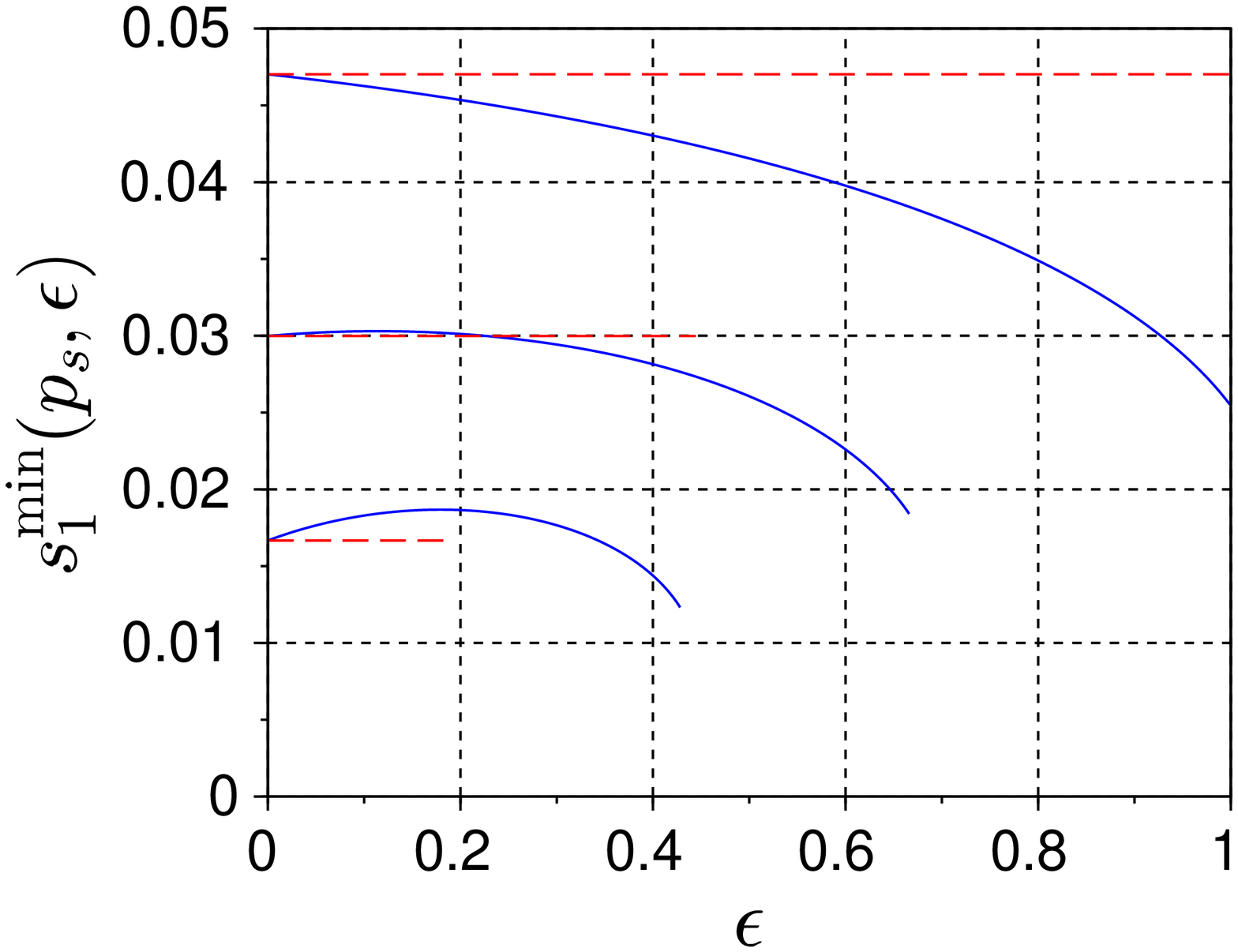}
		\caption{\textbf{(a)} Blue solid lines --- plots of $s_1^{\min}(p_s, \epsilon)$ versus $p_s$ varied from 0 to $p_{s\max}$ as per \eqref{eq_psmax}, at $\epsilon=0.1$, 0.5, 1 (from right to left). Red dashed line --- plot of the asymptotic formula \eqref{eq_mainresult}. \textbf{(b)} Blue solid lines --- plots of $s_1^{\min}(p_s, \epsilon)$ versus $\epsilon$ varied from 0 to $\epsilon_{\max}$ as per \eqref{eq_epmax}, at $p_s=0.5$, 0.6, 0.7 (from top to bottom). Vertical position of red dashed lines is the result of \eqref{eq_mainresult}, horizontal span denotes the estimated applicability range \eqref{eq_I0appx_condb}.}\label{fig_s1m}
\end{figure*}

Except for the degenerate cases \eqref{eq_degenerate}, $g(s_1)$ is negative at $s_1=0$
\beq\label{eq_gs10}
g(s_1=0)=-I_0(p_s, \epsilon)<0
\eeq
and tends to $+0$ \footnote{Notations $-0$ and $+0$ denote the left ang right one-sided limits.} with $s_1\to 1-0$ as soon as
\beq
\lim_{s_1\to 1-0} \; \frac{I_0\big((1-s_1)p_s, \epsilon \big)}{2 I_0\big((1-\sqrt{s_1})p_s, \epsilon \big)}=2,
\eeq
hence $g(s_1)$ changes sign at least once on $s_1\in(0,1)$. According to numerical evidence \footnote{This statement was confirmed up to machine precision for each combination of $p_s\in[0.01, 0.99]$ and $\rho\in[0.01, 1]$ (both with step 0.01).}, we assume that $g(s_1)$ changes sign exactly once on $(0,1)$ without providing a rigorous proof for the latter statement (note however that for the asymptotic case \eqref{eq_mainresult} this statement is rigorous). In line with the above, the solution to \eqref{eq_sufcond_a} has the form
\beq\label{eq_s1range}
s_1^{\min}(p_s, \epsilon) < s_1 <1,
\eeq
where $s_1^{\min}(p_s, \epsilon)$ is the unique root of $g(s_1)$ on $(0,1)$. Several plots of $s_1^{\min}(p_s, \epsilon)$ versus $p_s$ at $\epsilon$ fixed and versus $\epsilon$ at $p_s$ fixed, which are obtained by numerically solving for the zero of $g(s_1)$, are shown in Fig.~\ref{fig_s1m} with blue solid lines.

Further insight into the dependence of mutual information $I_{xy}$ (and, consequently, of $\Phi_{\text{eff}}$ and II) upon parameters can be obtained by inserting the expressions for the two-time probabilities \eqref{eq_psbeps} into the definition of $I_0(p_s,\epsilon)$ in \eqref{eq_I0} and expanding it in powers of $\epsilon$ (weak time correlation limit), which yields
\beq
I_0(p_s,\epsilon)=\frac{1}{2 \log 2} \left( \frac{p_s}{1-p_s} \right)^2 \cdot \epsilon^2 + O(\epsilon^3).
\eeq
Estimating the residual term (see details in Appendix~\ref{sec_I0appxderiv}) indicates that the approximation by the leading term
\beq\label{eq_I0appx}
I_0(p_s,\epsilon) \approx \frac{\epsilon^2}{2 \log 2} \left( \frac{p_s}{1-p_s} \right)^2
\eeq
is valid when
\begin{subequations}\label{eq_I0appx_cond}
	\begin{align}
		|\epsilon| &\ll 1,\label{eq_I0appx_conda} \\
		|\epsilon| &\ll \left(\frac{p_b}{p_s}\right)^2 \! = \epsilon_{\max}^2 \:. \label{eq_I0appx_condb}
	\end{align}
Solving \eqref{eq_I0appx_condb} for $p_s$ rewrites it in the form of an upper bound \footnote{The use of `$\ll$' sign is not appropriate in \eqref{eq_I0appx_condc}, because this inequality does not imply a small ratio between its left-hand and right-hand parts.} for $p_s$
\beq\label{eq_I0appx_condc}
	p_s < \frac{1}{1+\sqrt{|\epsilon|}}.
\eeq
\end{subequations}
Note how inequalities \eqref{eq_I0appx_condb}, \eqref{eq_I0appx_condc} compare to the formal upper bounds $\epsilon_{\max}$ in \eqref{eq_epmax} and $p_{s\max}$ in \eqref{eq_psmax} which arise from the definition of $\epsilon$ \eqref{eq_ep} due to the requirement of positive probabilities.

Approximation \eqref{eq_I0appx} is plotted in Fig.~\ref{fig_I0} with red dashed lines along with corresponding upper bounds of approximation applicability range \eqref{eq_I0appx_condc} denoted by red dots (note that large $\epsilon$ violates \eqref{eq_I0appx_conda} anyway, thus in this case \eqref{eq_I0appx_condc} has no effect). Mutual information \eqref{eq_I0appx} scales with $\epsilon$ within range \eqref{eq_I0appx_cond} as $\epsilon^2$ and vanishes with $\epsilon\to 0$. The same holds for effective information \eqref{eq_Ieff0}. Since the normalizing denominator in \eqref{eq_II_b} contains one-time entropies which do not depend on $\epsilon$ at all, this scaling of $\Phi_{\text{eff}}$ does not change the minimum information bipartition, finally implying that II also scales as $\epsilon^2$. That said, as factor $\epsilon^2$ does not affect the sign of $\Phi_{\text{eff}}$, the lower bound $s_1^{\min}$ in \eqref{eq_s1range} exists and is determined only by $p_s$ in this limit.

Substituting the approximation \eqref{eq_I0appx} for $I_0(\cdot)$ into the definition of $g(s_1)$ in \eqref{eq_sufcond_b} after simplifications reduces the equation $g(s_1)=0$ to the following \footnote{See comment below Eq.~\eqref{eq_IxyI0}.}:
\beq\label{eq_s1asymp}
p_s (\sqrt{2}-1)\: s_1 -\sqrt{s_1} + (1-p_s)(\sqrt{2}-1) =0\;,
\eeq
whose solution in terms of $s_1$ on $0<s_1<1$ equals $s_1^{\min}$, according to the reasoning behind Eq.~\eqref{eq_s1range}. Solving \eqref{eq_s1asymp} as a quadratic equation in terms of $\sqrt{s_1}$ produces a unique root on $(0,1)$, which yields
\beq\label{eq_mainresult}
s_1^{\min}(p_s)|_{\epsilon\to 0} = \left(
\frac{1-\sqrt{1-4p_s (1-p_s) (\sqrt{2}-1)^2}}
{2p_s (\sqrt{2}-1)}
\right)^2.
\eeq

Result of \eqref{eq_mainresult} is plotted in Fig.~\ref{fig_s1m} with red dashed lines: in panel (a) as a function of $p_s$, and in panel (b) as horizontal lines whose vertical position is the result of \eqref{eq_mainresult}, and horizontal span denotes the estimated applicability range \eqref{eq_I0appx_condb} (note that condition \eqref{eq_I0appx_conda} also applies, and becomes stronger than \eqref{eq_I0appx_condb} when $p_s<1/2$).

\section{Comparison of Integrated Information Measures}\label{sec_cmp}
In this Section we compare the outcome of two versions of empirical Integrated Information measures available in the literature, one being the ``all-minus-sum'' effective information \eqref{eq_Ieff} from \cite{barrett2011practical} which is used elsewhere in this study, and the other ``decoder based'' information as introduced in \cite{oizumi2016measuring} and expressed by Eqs.~(\ref{eq_Phistar}a-c). We calculate both measures by their respective definitions using the one- and two-time probabilities from Eqs.~(\ref{eq_Ponetime}a,b) and (\ref{eq_Ptwotime}a-d) for the spiking-bursting model with $N=6$ bits, assuming no spatial correlations among bits in spiking activity, with same spike probability $P$ in each bit. In this case
\beq
s_x=P^{m(x)} (1-P)^{N-m(x)},\quad P=s_1^{\frac{1}{N}},
\eeq
where $m(x)$ is the number of ones in the binary word $x$.

We consider only a symmetric bipartition with subsystems $A$ and $B$ consisting of $N/2=3$ bits each. Due to the assumed equal spike probabilities in all bits and in the absence of spatial correlations of spiking, this implies complete equivalence between the subsystems. In particular, in the notations of Sec.~\ref{sec_ii} we get
\beq
s_1=s_A s_B, \quad s_A=s_B=\sqrt{s_1}.
\eeq

This choice of the bipartition is firstly due to the fact that the sign of effective information for this bipartition determines the sign \footnote{Although the actual value of II is determined by the minimal information bipartition which may be different.} of the resultant ``whole minus sum'' II. This has been established in Sec.~\ref{sec_ii} (see reasoning behind Eqs.~\eqref{eq_condmin}--\eqref{eq_sufcond} and further on); moreover, this effective information has been denoted in Eq.~\eqref{eq_sufcond} as a function
\beq\label{eq_Phieff_gs1}
\Phi_{\text{eff}}(AB) = g(s_1),
\eeq
which has been analyzed in Sec.~\ref{sec_ii}.

Moreover, the choice of the symmetric bipartition is consistent with available comparative studies of II measures \cite{mediano2019measuring}, where it was substantiated by the conceptual requirement that highly asymmetric partitions should be excluded \cite{balduzzi2008integrated}, and by the lack of a generally accepted specification of minimum information bipartition; for further discussion, see \cite{mediano2019measuring}.

\begin{figure*}
	\centering
	\textbf{(a)}\includegraphics[width=0.46\textwidth]{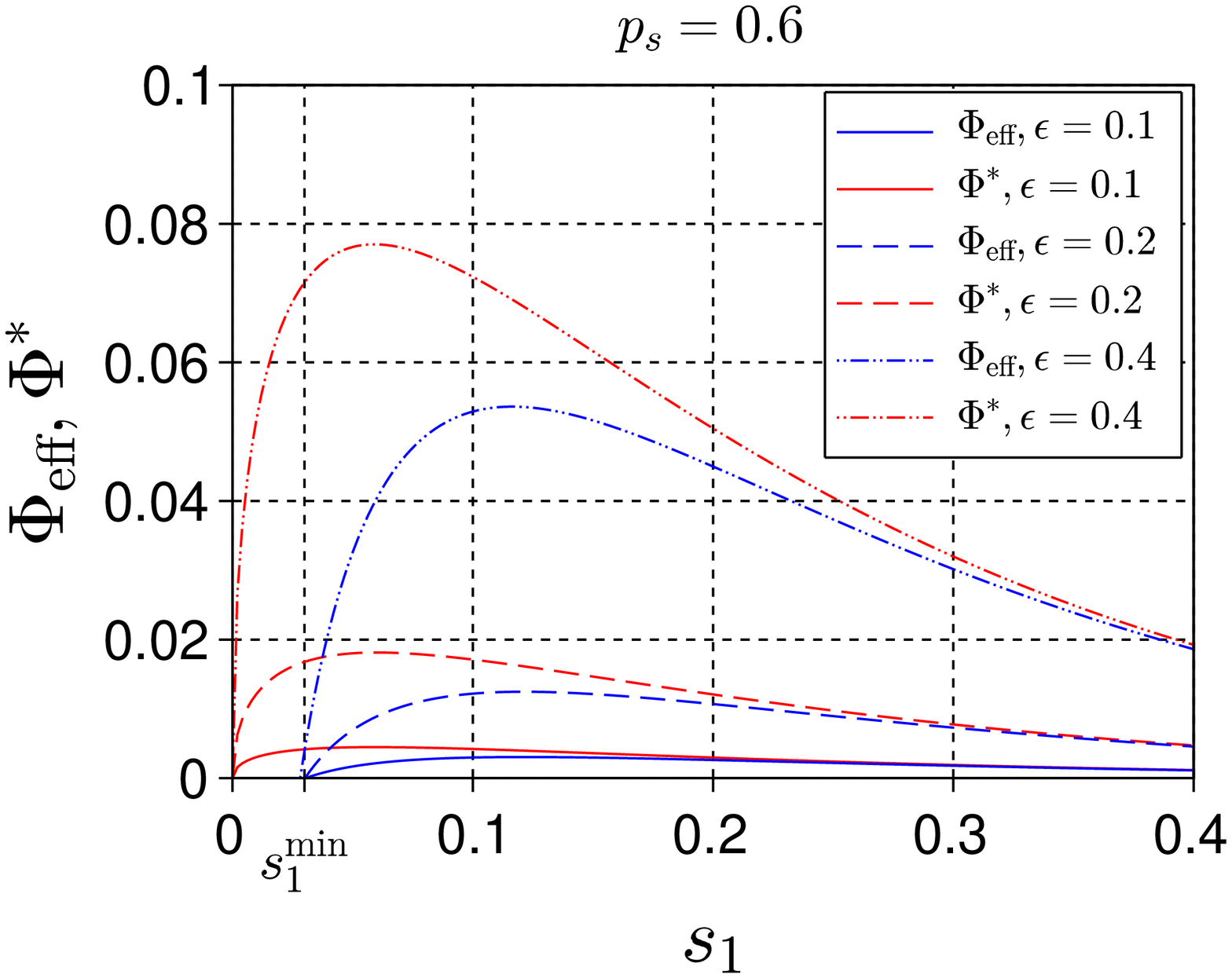}\hfill
	\textbf{(b)}\includegraphics[width=0.46\textwidth]{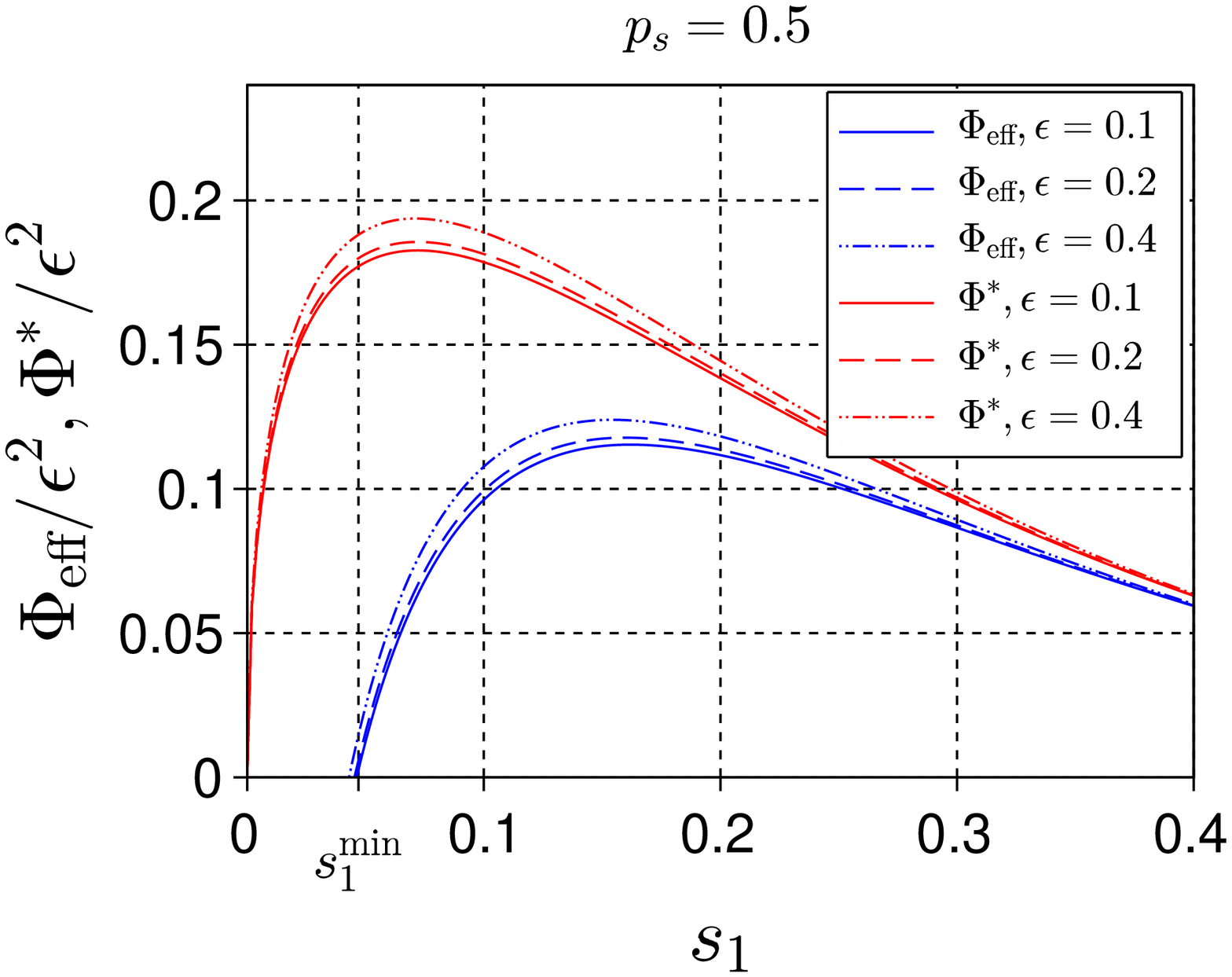}\\
	\textbf{(c)}\includegraphics[width=0.46\textwidth]{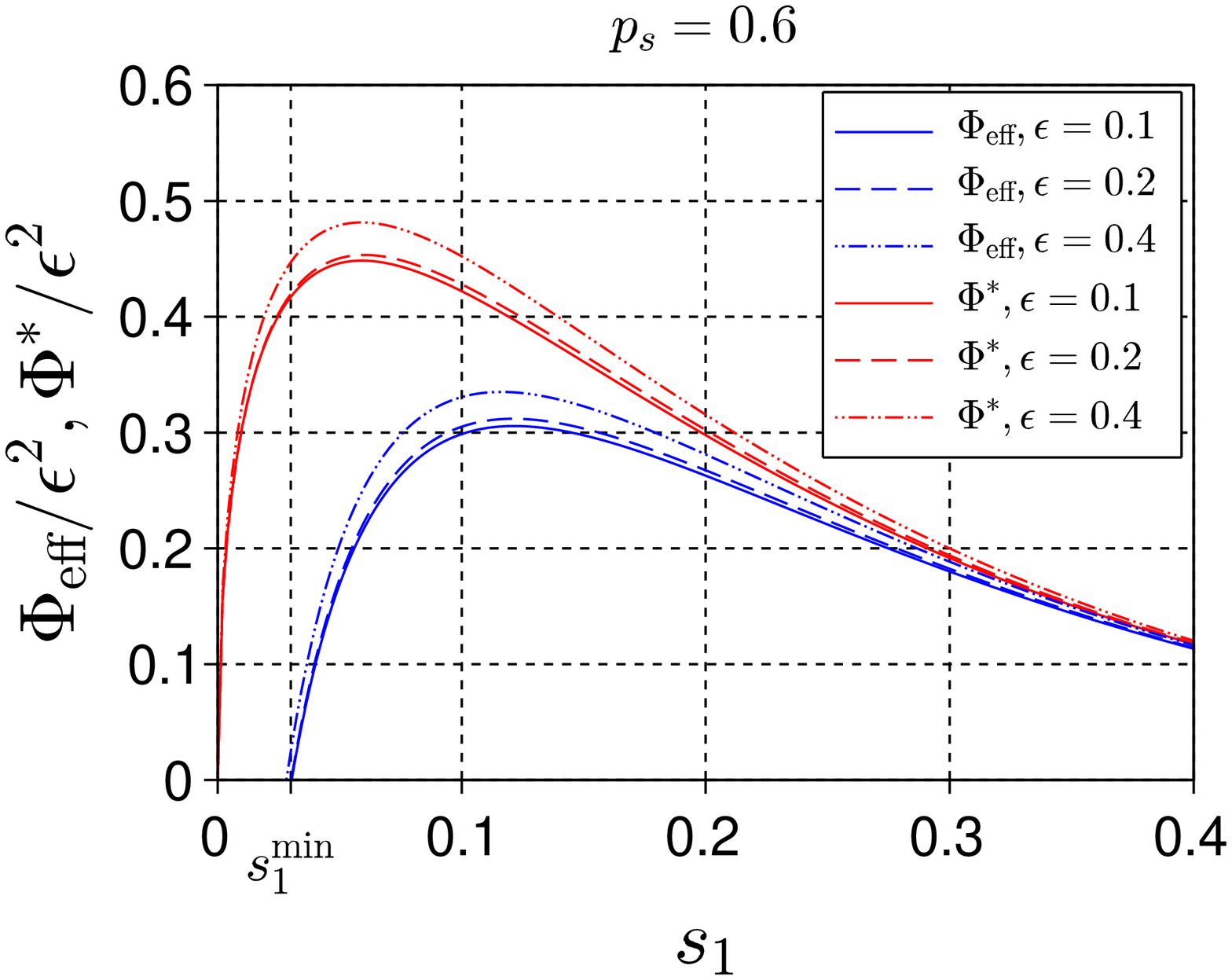}\hfill
	\textbf{(d)}\includegraphics[width=0.46\textwidth]{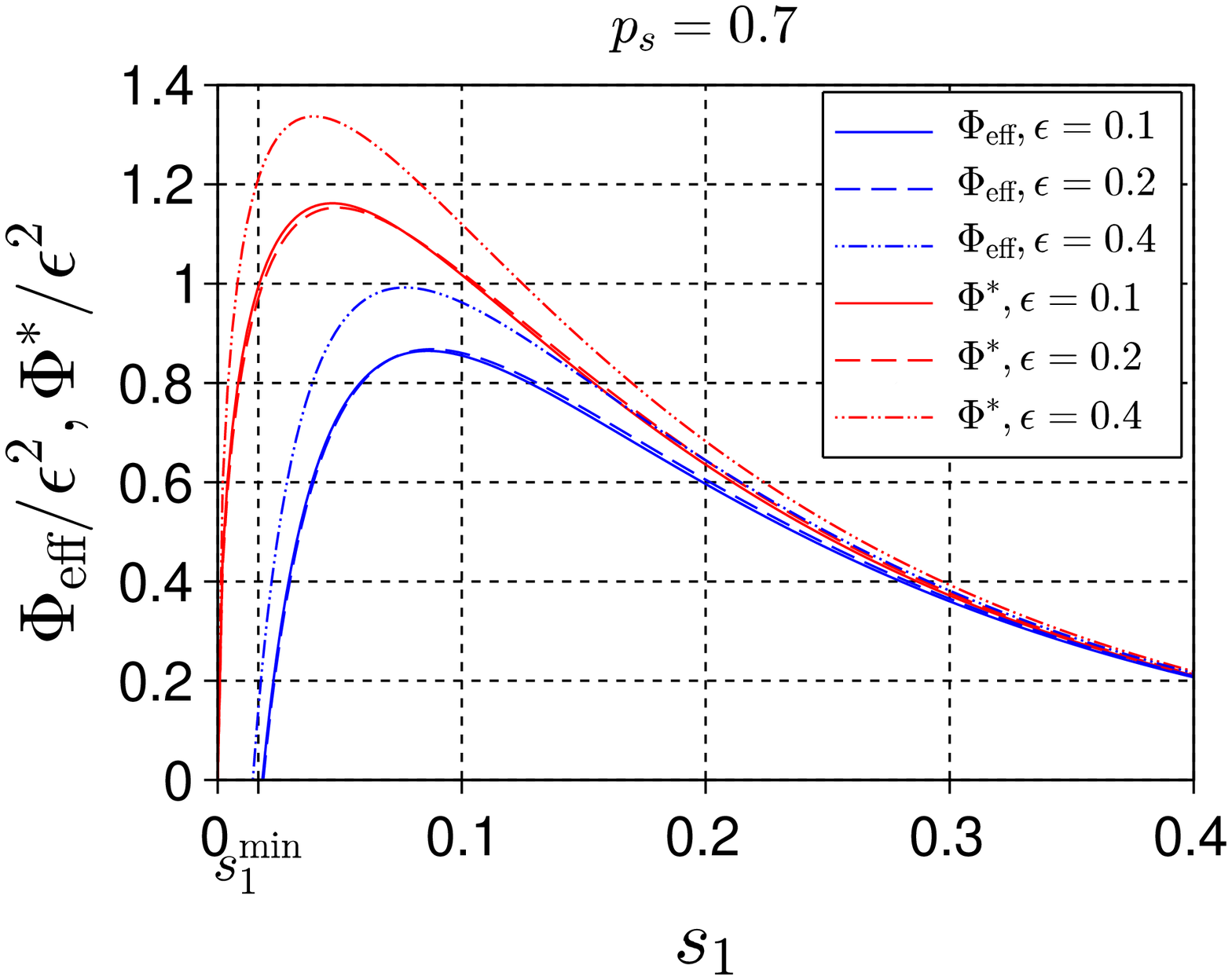}\\
		\caption{Comparison of two versions of empirical effective information for the symmetric bipartition --- ``whole-minus-sum'' measure \eqref{eq_Ieff} from \cite{barrett2011practical} (blue lines) and ``decoder based'' information \eqref{eq_Phistar} from \cite{oizumi2016measuring} (red lines) versus spiking activity parameter $s_1$ at various fixed values of the bursting component parameters $p_s$ (indicated on top of the panels) and $\epsilon$ (indicated in the legends). Panel \textbf{(a)} --- unnormalized values, panels \textbf{(b)}--\textbf{(d)} --- normalized by $\epsilon^2$. Threshold $s_1^{\min}$ calculated according to \eqref{eq_mainresult} is shown in each panel with an additional vertical grid line.}\label{fig_cmp}
\end{figure*}

We have studied the dependence of the mentioned effective information measures upon spiking activity, which is controlled by $s_1$, at different fixed values of the parameters $p_s$ and $\epsilon$ specifying the bursting component. Typical dependence of both measures upon $s_1$, taken at $p_s=0.6$ with several values of $\epsilon$, is shown in Fig.~\ref{fig_cmp}, panel (a).

The behavior of the ``whole minus sum'' effective information $\Phi_{\text{eff}}$ \eqref{eq_Phieff_gs1} (blue lines in Fig.~\ref{fig_cmp}) is found to agree with the analytical findings of Sec.~\ref{sec_ii}:
\begin{itemize}
	\item $\Phi_{\text{eff}}$ transitions from negative values to positive at a certain threshold value of $s_1=s_1^{\min}$, which is well approximated by the formula \eqref{eq_mainresult} when $\epsilon$ is small, as required by (\ref{eq_I0appx_cond}a,b); the result of Eq.~\eqref{eq_mainresult} is indicated in each panel of Fig.~\ref{fig_cmp} by an additional vertical grid line labeled $s_1^{\min}$ on the abscissae axis, cf.~Fig.~\ref{fig_s1m};
	\item $\Phi_{\text{eff}}$ reaches a maximum on the interval $s_1^{\min}<s_1<1$ and tends to zero (from above) at $s_1\to 1$;
	\item $\Phi_{\text{eff}}$ scales with $\epsilon$ as $\epsilon^2$, when (\ref{eq_I0appx_cond}a,b) hold.
\end{itemize}

To verify the scaling observation, we plot the scaled values of both information measures $\Phi_{\text{eff}}/\epsilon^2$, $\Phi^{\ast}/\epsilon^2$ in the panels (b)--(d) of Fig.~\ref{fig_cmp} for several fixed values of $p_s$ and $\epsilon$. Expectedly, the scaling fails at $p_s=0.7$, $\epsilon=0.4$ in panel (d), as (\ref{eq_I0appx_cond}b) is not fulfilled in this case.

Furthermore, the ``decoder based'' information $\Phi^{\ast}$ (plotted with red lines in Fig.~\ref{fig_cmp}) behaves mostly the same way, apart from being always non-negative (which was one of key motivations for introducing this measure in \cite{oizumi2016measuring}). At the same time, the sign transition point $s_1^{\min}$ of the ``whole minus sum'' information associates with a rapid growth of the ``decoder based'' information. When $s_1$ is increased towards 1, the two measures converge. Remarkably, the scaling as $\epsilon^2$ is found to be shared by both effective information measures.

\section{Discussion}

In general, the spiking-bursting model is completely specified by the combination of a full single-time probability table $s_x$ (consisting of $2^N$ probabilities of all possible outcomes, where $N$ is the number of bits) for the time-uncorrelated spontaneous activity, along with two independent parameters (e.g. $p_s$ and $\epsilon$) for the dichotomous component. This combination is, however, redundant in that it admits a one-parameter scaling \eqref{eq_scaling} which leaves the resultant stochastic process invariant.

Condition \eqref{eq_sufcond} was derived assuming that spiking activity in individual bits (i.e. nodes, or neurons) constituting the system is independent among the bits, which implies that the probability table $s_x$ is fully determined by $N$ spike probabilities for individual nodes. The condition is formulated in terms of $p_s$, $\epsilon$ and a single parameter $s_1$ (system-wide spike probability) for the spontaneous activity, thus ignoring the ``internal structure'' of the system, i.e. the spike probabilities for individual nodes. This condition provides that the ``whole minus sum'' effective information is positive for any bipartition, regardless of the mentioned internal structure. Moreover, in the limit \eqref{eq_I0appx_cond} of weak correlations in time, the inequality \eqref{eq_sufcond_a} can be explicitly solved in terms of $s_1$, producing the solution \eqref{eq_s1range}, \eqref{eq_mainresult}.

In this way, the inequality \eqref{eq_s1range} together with the asymptotic estimate \eqref{eq_mainresult} supplemented by its applicability range \eqref{eq_I0appx_cond} specifies the region in the parameter space of the system, where the ``whole minus sum'' II is positive regardless of the internal system structure (\emph{sufficient} condition). The internal structure (though still without spike correlations across the system) is taken into account by the \emph{necessary and sufficient} condition \eqref{eq_condmin} for positive II.

The mentioned conditions were derived under the assumption of absent correlation between spontaneous activity in individual bits \eqref{eq_sAB}. If correlation exists and is \emph{positive}, then $s_1> s_A s_B$, or $s_B < s_1/s_A$. Then comparing the expressions for $\Phi_{\text{eff}}$ \eqref{eq_Ieff0} (general case) to \eqref{eq_fsdef} (space-uncorrelated case), and taking into account that $I_0(p_s)$ is an increasing function, we find $\Phi_{\text{eff}}<f(s_A)$, cf. \eqref{eq_fsdef_a}. This implies that any \emph{necessary} condition for positive II remains as such. Likewise, in the case of \emph{negative} correlations we get $\Phi_{\text{eff}}>f(s_A)$, implying that a \emph{sufficient} condition remains as such.

We found that II scales as $\epsilon^2$ for $\epsilon$ small (namely, within \eqref{eq_I0appx_cond}) when other parameters (i.e. $p_s$ and spiking probability table $s_x$) are fixed. For the ``whole minus sum'' information, this is an analytical result. Note that the reasoning behind this result does not rely upon the assumption of spatial uncorrelation of spiking activity (between bits) and thus applies to arbitrary spiking-bursting systems. According to a numerical calculation, this scaling is applicable to the ``decoder based'' information as well.

Remarkably, II can not exceed the time delayed mutual information for the system as a whole, which in case of the spiking-bursting model in its present formulation is no greater than 1 bit.

The present study substantiates, refines and quantifies qualitative observations in regard to II in the spiking-bursting model which were initially made in \cite{AstroNeur}. The existence of lower bounds in spiking activity (characterized by $s_1$) which was noticed in \cite{AstroNeur} is now expressed in the form of an explicit inequality \eqref{eq_s1range} with the estimate \eqref{eq_mainresult} for the bound $s_1^{\min}$. The observation of \cite{AstroNeur} that typically $s_1^{\min}$ is mostly determined by burst probability and weakly depends upon time correlations of bursts also becomes supported by the quantitative result \eqref{eq_s1range}, \eqref{eq_mainresult}.

The model provides a basis for possible modifications in order to apply Integrated Information concepts to systems exhibiting similar, but more complicated behavior (in particular, to neuron-astrocyte networks). Such modifications might incorporate non-trivial spatial patterns in bursting, and causal interactions within and between the spiking and bursting subsystems.

The model can also be of interest as a new discrete-state test bench for different formalizations of Integrated Information, while available comparative studies of II measures mainly focus on Gaussian autoregressive models \cite{mediano2019measuring, tegmark2016improved}.

\section*{Acknowledgments}
This research was funded by the Ministry of Education and Science of the Russian Federation within Agreement No.~075-15-2019-871.

\appendix

\section{Derivation of parameters scaling of the spiking-bursting model}\label{sec_scaleder}

In order to formalize the reasoning in Section~\ref{sec_scaling}, we introduce an auxiliary 3-state process $W$ with set of one-time states $\{s', d, b \}$, where $s'$ and $b$ are always interpreted as spiking and bursting states in terms of Section~\ref{sec_model}, and $d$ is another state, which is assumed to produce all bits equal 1 like in a burst, but in a time-uncorrelated manner (which is formalized by Eq.~\eqref{eq_Wcond2} below) like in a system-wide spike. When $W$ is properly defined (by specifying all necessary probabilities, see below) and supplemented with a time-uncorrelated process $S$ as a source of spontaneous activity for the state $s'$, these together constitute a completely defined stochastic model $\{W,S\}$.

This 3-state based model may be mapped on equivalent (in terms of resultant realizations) 2-state based models as in Section~\ref{sec_model} in an ambiguous way, because the state $d$ may be equally interpreted either as a system-wide spike, or as a time-uncorrelated burst, thus producing two different dichotomous processes (which we denote as $V$ and $V'$) for the equivalent spiking-bursting models. The relationship between the states of $W$, $V$ and $V'$ is illustrated by the following diagram.
\beq\label{eq_diagram}
\begin{array}{rccc}
	V=  & \multicolumn{2}{c}{s} & b\\
	& \multicolumn{2}{c}{\overbrace{\vphantom{x}\hfill}} & \\[-2ex]
	W=  & s' & d & b \\[-2ex]
	&  & \multicolumn{2}{c}{\underbrace{\vphantom{x}\hfill}} \\
	V'= & s' & \multicolumn{2}{c}{b'}
\end{array}
\eeq

As soon as $d$-states of $W$ are interpreted in $V$ as (spiking) $s$-states, the spontaneous activity process $S$ accompanying $V$ has to be supplemented with system-wide spikes whenever $W=d$, in addition to the spontaneous activity process $S'$ for $V'$. In order to maintain the absence of time correlations in spontaneous activity (which is essential for the analysis in Section~\ref{sec_ii}), we assume time-uncorrelated choice between $W=s'$ and $W=d$ when $V=s$ (which manifests below in Eq.~\eqref{eq_Wcond2}). Then the difference between the spontaneous components $S$ and $S'$ comes down to a difference in the corresponding one-time probability tables $s_x$ and $s'_x$.

In the following, we proceed from the dichotomous process $V$ defined as in Section~\ref{sec_model}, then define a consistent 3-state process $W$, and further obtain another dichotomous process $V'$ for an equivalent model. Finally, we establish the relation between the corresponding probability tables of spontaneous activity $s_x$ and $s'_x$.

The first dichotomous process $V$ has states denoted by $\{s, b \}$ and is related to $W$ according to the rule $V=s$ when $W=s'$ or $W=d$, and $V=b$ whenever $W=b$ (see diagram \eqref{eq_diagram}). Assume fixed conditional probabilities
\begin{subequations}\label{eq_Wcond}
	\begin{align}
		p(W=s' \;|\; V=s) &= \alpha,\\
		p(W=d \;|\; V=s) &= \beta=1-\alpha,
	\end{align}
\end{subequations}
which implies one-time probabilities for $W$ as
\beq\label{eq_Wab}
p_{s'}=\alpha p_s,\quad p_{d}=\beta p_s.
\eeq

The mentioned requirement of time-uncorrelated choice between $W=s'$ and $W=d$ when $V=s$ is expressed by factorized two-time conditional probabilities
\begin{subequations}\label{eq_Wcond2}
	\begin{align}
		p(W=s's' \;|\; V=ss) &= \alpha^2,\\
		p(W=s'd \;|\; V=ss) &=\alpha \beta = p(W=ds' \;|\; V=ss),\\
		p(W=dd \;|\; V=ss) &= \beta^2.
	\end{align}
\end{subequations}
Given the two-time probability table for $V$ \eqref{eq_psbmat} along with  the conditional probabilities \eqref{eq_Wcond}, \eqref{eq_Wcond2}, we arrive at a two-time probability table for $W$ 
\beq\label{eq_W}
p_{q\in\{s', d, b \},r\in\{s', d, b \}}= 
\begin{blockarray}{lccc}
	& s' &  d & b \\
	\begin{block}{l(cc|c)}
		s' & \alpha^2 p_{ss} & \alpha \beta p_{ss} & \alpha p_{sb} \\
		d & \alpha \beta p_{ss} & \beta^2 p_{ss} & \beta p_{sb} \\ \BAhhline{~---}
		b & \alpha p_{bs} & \beta p_{bs} & p_{bb}\\
	\end{block}
\end{blockarray}.
\eeq
Note that \eqref{eq_W} is consistent both with \eqref{eq_Wab}, which is obtained by summation along the rows of \eqref{eq_W}, and with \eqref{eq_psbmat}, which is obtained by summation within the line-separated cell groups in \eqref{eq_W}:
\begin{subequations}
	\begin{align}
		p_{ss}& \equiv p_{s's'}+p_{s'd}+p_{ds'}+p_{dd}\\
		p_{sb}& \equiv p_{s'b}+p_{db}\\
		p_{bs}& \equiv p_{bs'}+p_{bd}\\
		p_{bb}& \equiv p_{bb}.
	\end{align}
\end{subequations}

Consider the other dichotomous process $V'$ with states $\{s', b' \}$ obtained from $W$ according to the rule $V'=b'$ when $W=d$ or $W=b$, and $V'=s'$ whenever $W=s'$ (see diagram \eqref{eq_diagram}). The two-time probability table for $V'$ is obtained by another partitioning of the table \eqref{eq_W}
\beq
p_{qr}=
\begin{blockarray}{lccc}
	& s' &  d & b \\
	\begin{block}{l(c|cc)}
		s' & \alpha^2 p_{ss} & \alpha \beta p_{ss} & \alpha p_{sb} \\
		\BAhhline{~---}
		d & \alpha \beta p_{ss} & \beta^2 p_{ss} & \beta p_{sb} \\ 
		b & \alpha p_{bs} & \beta p_{bs} & p_{bb}\\
	\end{block}
\end{blockarray}
\eeq
with subsequent summation of cells within groups, which yields
\begin{subequations}\label{eq_Vprime2}
	\begin{align}
		p_{s's'} &= \alpha^2 p_{ss},\\
		p_{s'b'} &= \alpha(\beta p_{ss}+p_{sb}) = p_{b's'},\\
		p_{b'b'} &= \beta^2 p_{ss} + 2 \beta p_{sb} + p_{bb}.
	\end{align}
\end{subequations}
The corresponding one-time probabilities for $V'$ read
\begin{subequations}\label{eq_Vprime1}
	\begin{align}
		p_{s'}&=\alpha p_s,\\
		p_{b'}&=\beta p_s + p_b.
	\end{align}
\end{subequations}

In order to establish the relation between the one-time probability tables of spontaneous activity $s_x$ and $s'_x$, we equate the resultant one-time probabilities of observing a given state $x$ as per \eqref{eq_Ponetime} for the two equivalent models $\{V,S\}$ and $\{V',S'\}$
\begin{subequations}
	\begin{align}
		p(x\neq 1) &= p_s s_x = p_{s'} s'_x, \\
		p(x=1) &= p_s s_1 + p_b = p_{s'} s'_1 + p_{b'}.
	\end{align}
\end{subequations}
Taking into account \eqref{eq_Vprime1}, we finally get
\begin{subequations}\label{eq_scalesx}
	\begin{align}
		s_x &= \alpha s'_x, \\
		1-s_1 &= \alpha (1- s'_1).
	\end{align}
\end{subequations}

Equations \eqref{eq_Vprime2}, \eqref{eq_Vprime1} and (\ref{eq_scalesx}) fully describe the transformation of the spiking-bursting model which keeps the resultant stochastic process invariant by the construction of the transform. Taking into account that the dichotomous process is fully described by just two independent quantities, e.g. $p_s$ and $p_{ss}$, all other probabilities being expressed in terms of these due to normalization and stationarity, the full invariant transformation is uniquely identified by a combination of (\ref{eq_scalesx}a,b), (\ref{eq_Vprime2}a) and (\ref{eq_Vprime1}a), which together constitute the scaling~\eqref{eq_scaling}.

Note that parameter $\alpha$ within its initial meaning \eqref{eq_Wcond} may take on values in the range $0<\alpha\le 1$ (case $\alpha=1$ producing the identical transform). That said, in terms of the scaling (\ref{eq_scaling}a-d), all values $\alpha>0$ are equally possible, so that mutually inverse values $\alpha=\alpha_1$ and $\alpha=\alpha_2=1/\alpha_1$ produce mutually inverse transforms.

\section{Expressing mutual information for the spiking-bursting process}\label{sec_deriv}
One-time entropy $H_x$ for the spiking-bursting process is expressed by 
\eqref{eq_Hx} with probabilities $p(x)$ taken from \eqref{eq_Ponetime}:
\beq
H_x=\sum_x \{p(x)\}=\sum_x \{p_s s_x\} + \{p_1\} - \{p_s s_1\},
\eeq
where the additional terms besides the sum over $x$ account for the specific expression \eqref{eq_Ponetime_p1} for $p(x=1)$. Using the relation
\beq\label{eq_bracerule}
\{ab\} \equiv a\{b\}+\{a\}b,
\eeq
which is derived directly from \eqref{eq_braces}, and collecting similar terms, we arrive at
\beq\label{eq_exprHx}
H_x=p_s H_s - p_s \{s_1\} + (1-s_1)\{p_s\} + \{p_1\},
\eeq
where $H_s$ is the entropy of the spiking component taken alone
\beq\label{eq_Hs}
H_s=\sum_x \{s_x\}.
\eeq

Two-time entropy is expressed similarly, by substituting probabilities $p(xy)$ from \eqref{eq_Ptwotime} into the definition of entropy and taking into account the special cases with $x=1$ and/or $y=1$:
\begin{widetext}
\begin{equation}\label{eq_Hxy_pre}
\begin{split}
	H_{xy}=\sum_{xy} \{p(xy)\} = \sum_{xy} \{p_{ss} s_x s_y\}
	&-\sum_x \{p_{ss} s_x s_1\} + \sum_x \{\pi s_x\}\\
	&-\sum_y \{p_{ss} s_1 s_y\} + \sum_y \{\pi s_y\}
	+\{p_{ss} s_1^2\} - 2\{\pi s_1\} + \{p_{11}\}.
\end{split}
\end{equation}
\end{widetext}
Further, applying \eqref{eq_bracerule} and using the notation \eqref{eq_Hs}, we find
\begin{subequations}\label{eq_simpl}
\beq\label{eq_simpl1}
\begin{split}
		\sum_{xy} \{p_{ss} s_x s_y\} &= p_{ss} \sum_{xy} \{s_x s_y\} + \{p_{ss} \} \sum_{xy} s_x s_y \\
		&=p_{ss}\cdot 2H_s + \{p_{ss} \},
\end{split}
\eeq
where we used the reasoning that $\sum_{xy} \{s_x s_y\}$ is the two-time entropy of the spiking component taken alone, which is (due to the postulated absence of time correlations in it) twice the one-time entropy $H_s$ (this of course can equally be found by direct calculation). Similarly, we get
\beq\label{eq_simpl2}
\begin{split}
	\sum_{x} \{p_{ss} s_x s_1\} &= p_{ss} s_1 \sum_{x} \{s_x\}
	+ \{p_{ss} s_1\} \sum_{x} s_x\\
	&= p_{ss} s_1 H_s + \{p_{ss} s_1\}
\end{split}
\eeq
and exactly the same expression for $\sum_y \{p_{ss} s_1 s_y\}$, and also
\beq\label{eq_simpl3}
\begin{split}
	\sum_y \{\pi s_y\} = \sum_x \{\pi s_x\} &= \pi \sum_x \{s_x\}
	+ \{\pi\} \sum_x s_x\\
	&= \pi H_s + \{\pi\}.
\end{split}
\eeq
\end{subequations}

Substituting (\ref{eq_simpl}a-c) into \eqref{eq_Hxy_pre}, using \eqref{eq_bracerule} where applicable, and collecting similar terms with the relation
\beq
p_{ss}+\pi-p_{ss}s_1 \equiv p_s
\eeq
taken into account, we arrive at
\begin{multline}\label{eq_exprHxy}
H_{xy}=2p_s H_s + (1-s_1)^2 \{p_{ss}\} - 2 p_s \{s_1\} \\
+ 2(1-s_1)\{\pi\} + \{p_{11}\}.
\end{multline}

Finally, the expression \eqref{eq_Ixy} for mutual information is obtained by inserting \eqref{eq_exprHx} and \eqref{eq_exprHxy} into the definition \eqref{eq_defIxy}, with stationarity $H_y=H_x$ taken into account.

\section{Expanding $I_0$ in powers of $\epsilon$}\label{sec_I0appxderiv}
Taylor series expansion for a function $f(x)$ up to the quadratic term reads
\beq\label{eq_fxTayl}
f(x_0+\xi) = f(x_0) + f'(x_0) \xi + f''(x_0) \frac{\xi^2}{2} + R(\xi).
\eeq
The remainder term $R(\xi)$ can be represented in the Lagrange's form as
\beq
R(\xi) = f'''(c) \frac{\xi^3}{6},
\eeq
where $c$ is an unknown real quantity between $x_0$ and $x_0+\xi$.

The function $f(x)$ can be approximated by omitting $R(\xi)$ in \eqref{eq_fxTayl} if $R(\xi)$ is negligible compared to the quadratic term, for which it is sufficient that
\begin{subequations}\label{eq_Taylcond}
\beq\label{eq_Taylcond_a}
\left| f'''(c) \frac{\xi^3}{6} \right| \ll \left|f''(x_0) \frac{\xi^2}{2}\right|
\eeq
for any $c$ between $x_0$ and $x_0+\xi$, namely for
\beq\label{eq_Taylcond_b}
c \in
 \begin{cases}
 	(x_0, x_0 + \xi), & \text{if $\xi>0$},\\
 	(x_0 - |\xi|, x_0), & \text{if $\xi<0$}.
 \end{cases}
\eeq
\end{subequations}

Consider the specific case
\beq\label{eq_fx}
f(x)=-x \log x, \quad x>0,
\eeq
for which we get
\beq
f'(x)=-\log x -1,\quad f''(x)=-\frac{1}{x}, \quad f'''(x)=\frac{1}{x^2}.
\eeq

As long as $f'''(x)$ is a falling function for any $x>0$, fulfilling \eqref{eq_Taylcond_a} at the left boundary of \eqref{eq_Taylcond_b} (at $c=x_0$ if $\xi>0$, and at $c=x_0-|\xi|$ if $\xi<0$) makes sure \eqref{eq_Taylcond_a} is fulfilled in the whole interval \eqref{eq_Taylcond_b}. Precisely, the requirement is
\begin{subequations}
	\begin{align}
\left| \frac{1}{x_0^2} \frac{\xi^3}{6} \right| &\ll \left| \frac{1}{x_0}  \frac{\xi^2}{2} \right|, \quad \text{if} \quad \xi>0,\\
\left|\frac{1}{(x_0-|\xi|)^2} \frac{\xi^3}{6}\right| &\ll \left| \frac{1}{x_0}  \frac{\xi^2}{2} \right|, \quad \text{if} \quad \xi<0,
\end{align}
\end{subequations}
which in the case $\xi>0$ reduces to
\beq\label{eq_xipos}
\frac{\xi}{3 x_0} \ll  1,
\eeq
and in the case $\xi<0$ to
\begin{subequations}
\beq\label{eq_xineg}
\frac{1}{3}\Phi \left( \frac{|\xi|}{x_0} \right) \ll  1,
\eeq
where
\beq
\Phi(\zeta)=\frac{\zeta}{(1-\zeta)^2}.
\eeq
\end{subequations}
Replacing $\Phi(\cdot)$ in \eqref{eq_xineg} by its linearization $\Phi(\zeta)\approx\zeta$ for small $\zeta$, we reduce both \eqref{eq_xipos}  and \eqref{eq_xineg} to a single condition
\beq\label{eq_Taylcond2}
|\xi| \ll 3x_0.
\eeq

We use these considerations to expand in powers of $\epsilon$ the function $I_0(p_s,\epsilon)$ defined in \eqref{eq_I0} with $p_{ss}$, $p_{sb}$, $p_{bb}$ substituted by their expressions in terms of $\epsilon$ according to \eqref{eq_psbeps}. We note that the braces notation $\{\cdot\}$ defined in \eqref{eq_braces} is expressed via the function $f(x)$ from \eqref{eq_fx} as
\beq
\{q\}=\frac{f(q)}{\log 2}.
\eeq

Expanding this way the subexpressions of \eqref{eq_I0}
\begin{subequations}\label{eq_subexpr}
\begin{align}
\{p_{ss}\} &= \{p_s^2 +\epsilon p_s^2\},\\
\{p_{sb}\} &= \{p_s p_b -\epsilon p_s^2\},\\
\{p_{bb}\} &= \{p_b^2 +\epsilon p_s^2\},
\end{align}
\end{subequations}
we find by immediate calculation that the zero-order and linear in $\epsilon$ terms vanish, and the quadratic term yields \eqref{eq_I0appx}. The condition \eqref{eq_Taylcond2} has to be applied to all three subexpressions (\ref{eq_subexpr}a-c). Omitting the insignificant factor 3 in \eqref{eq_Taylcond2}, we obtain the applicability conditions
\begin{subequations}
	\begin{align}
	|\epsilon p_s^2| & \ll p_s^2,\\
	|\epsilon p_s^2| & \ll p_s p_b,\\
	|\epsilon p_s^2| & \ll p_b^2,
	\end{align}
\end{subequations}
which is equivalent to
\begin{subequations}\label{eq_Taylcond3}
	\begin{align}
	|\epsilon| & \ll 1,\\
	|\epsilon| & \ll \frac{p_b}{p_s} = \epsilon_{\max},\\
	|\epsilon| & \ll  \epsilon_{\max}^2,
	\end{align}
\end{subequations}
where the notation $\epsilon_{\max}$ from \eqref{eq_epmax} is used. We note that when $\epsilon_{\max}<1$, the condition (\ref{eq_Taylcond3}c) is the strongest among (\ref{eq_Taylcond3}a-c); when $\epsilon_{\max}>1$, the condition (\ref{eq_Taylcond3}a) is the strongest. Therefore, in both cases (\ref{eq_Taylcond3}b) can be dropped, thus producing \eqref{eq_I0appx_cond}.

\bibliographystyle{apsrev4-1}

\bibliography{refs}

\end{document}